\documentclass[aps,pra,reprint,superscriptaddress,amsmath,amssymb]{revtex4-2}

\usepackage{graphicx}
\usepackage{listings}
\usepackage{longtable}
\usepackage{color,soul}
\usepackage{algorithm}
\usepackage[bookmarks,bookmarksnumbered,colorlinks,allcolors=blue]{hyperref}
\usepackage{bookmark}
\usepackage{makecell}
\usepackage{csquotes}
\usepackage{amsmath}

\begin{document}

\title{Epitaxial titanium nitride microwave resonators: Structural, chemical, electrical, and microwave properties}

\author{Ran Gao}
\email{gaor410@gmail.com}
\affiliation{Alibaba Quantum Laboratory, Alibaba Group, Hangzhou, Zhejiang 311121, P.R.China}

\author{Wenlong Yu}
\affiliation{Alibaba Quantum Laboratory, Alibaba Group, Hangzhou, Zhejiang 311121, P.R.China}

\author{Hao Deng}
\affiliation{Alibaba Quantum Laboratory, Alibaba Group, Hangzhou, Zhejiang 311121, P.R.China}

\author{Hsiang-Sheng Ku}
\affiliation{Alibaba Quantum Laboratory, Alibaba Group, Hangzhou, Zhejiang 311121, P.R.China}

\author{Zhisheng Li}
\affiliation{Alibaba Quantum Laboratory, Alibaba Group, Hangzhou, Zhejiang 311121, P.R.China}

\author{Minghua Wang}
\affiliation{Westlake Center for Micro/Nano Fabrication, Westlake University, Hangzhou, Zhejiang 310024, P.R.China}

\author{Xiaohe Miao}
\affiliation{Instrumentation and Service Center for Physical Sciences, Westlake University, Hangzhou, Zhejiang 310024, P.R.China}

\author{Yue Lin}
\affiliation{Hefei National Laboratory for Physical Sciences at the Microscale, University of Science and Technology of China, Hefei, Anhui 230026, P.R.China}

\author{Chunqing Deng}
\email{dengchunqing@gmail.com}
\affiliation{Alibaba Quantum Laboratory, Alibaba Group, Hangzhou, Zhejiang 311121, P.R.China}

\begin{abstract}
    Titanium nitride is an attractive material for a range of superconducting quantum-circuit applications owing to its low microwave losses, high surface inductance, and chemical stability. The physical properties and device performance, nevertheless, depend strongly on the quality of the materials. In this paper, we focus on the highly crystalline and epitaxial titanium nitride thin films deposited on sapphire substrates using magnetron sputtering at an intermediate temperature (300$^{\circ}$C). We performed a set of systematic and comprehensive material characterizations to thoroughly understand the structural, chemical, and transport properties. Microwave losses at low temperatures were studied using patterned microwave resonators, where the best internal quality factor in the single-photon regime is measured to be $3.3\times 10^6$ and that in the high-power regime $> 1.0\times 10^7$. Adjusted with the material filling factor of the resonators, the microwave loss-tangent here compares well with the previously reported best values for superconducting resonators. This work lays the foundation of using epitaxial titanium nitride for low-loss superconducting quantum circuits.
\end{abstract}
\maketitle
\bookmarksetup{startatroot}

\section{Introduction}

The past decade has witnessed rapid progress in the improvement of coherence times for superconducting quantum circuits~\cite{Devoret2013, Kjaergaard2020}. Such advancement has largely benefited from the on-going quest for superior superconducting materials and the optimization of fabrication processes~\cite{Oliver2013}. Among the promising material candidates, titanium nitride (TiN) has long been studied and proven to have low microwave losses and excellent chemical stability for advanced quantum circuit fabrication \cite{Vissers2010,Diener2012,Chang2013,Melville2020,Krockenberger2012,Peltonen2018,Driessen2012}. Quality factors greater than $1\times 10^6$ have been reported for superconducting resonators made with TiN thin films grown using either reactive sputtering or molecular-beam epitaxy \cite{Vissers2010,Ohya2014,Richardson2020}. Factors in the fabrication processes such as etchant gas mixtures, surface-cleaning procedures, and etching profiles were also thoroughly studied and claimed to be critical to device performance~\cite{Sage2011,Sandberg2012,Bruno2015,Calusine2018,Lock2019}. Material properties, including stoichiometry, textures, and residual stress, have also been examined, but not to a great extent, to correlate with the microwave properties of superconducting resonators~\cite{Ohya2014}. 

Nevertheless, most of the studied TiN resonator devices are made of polycrystalline films, and few studies focus on epitaxial TiN superconducting resonators~\cite{Richardson2020}. Although the crystallinity and lattice structures of TiN films have not yet been shown to have a direct impact on the quality factors of microwave resonators, polycrystalline systems typically make it challenging to establish a comprehensive correlation between the material and microwave properties due to the pervasive existence of randomly distributed grains and grain boundaries. In particular, as the quality of the interfaces between the films and substrates have been shown to have substantial impacts on the coherence times of superconducting qubits~\cite{Wang2015, Dunsworth2017, McRae2020}, the incoherent interfaces in the polycrystalline heterostructures, as result of the non-uniform grains, hamper quantitative analysis and deterministic optimization at these regions. As such, the synthesis of a high-quality material system, \textit{i.e.,} epitaxial with high degree of crystallinity, is instrumental for in-depth studies on its physical and decoherence properties~\cite{Saveskul2019}. 

In addition, although there have been numerous implementations of high-quality epitaxial TiN films for various purposes, the deposition of the materials typically requires elevated deposition temperatures (\textit{i.e.,} 600$^{\circ}$C)-800$^{\circ}$C) or advanced material-synthesis systems such as molecular-beam epitaxy and pulsed-laser depositions \cite{Naik2014,Olson:2015,Krockenberger2012,Torgovkin2018,Rasic2017,Gupta2019,Saveskul2019,Guo2019}. These approaches are ideal for experimental research, yet mature for large-scale production and integration of superconducting quantum circuits. Considering process compatibility in device integration, it is advantageous to apply deposition techniques with low thermal budget and scale-up capabilities.

\begin{figure*} [t]
\includegraphics[width=12.9cm]{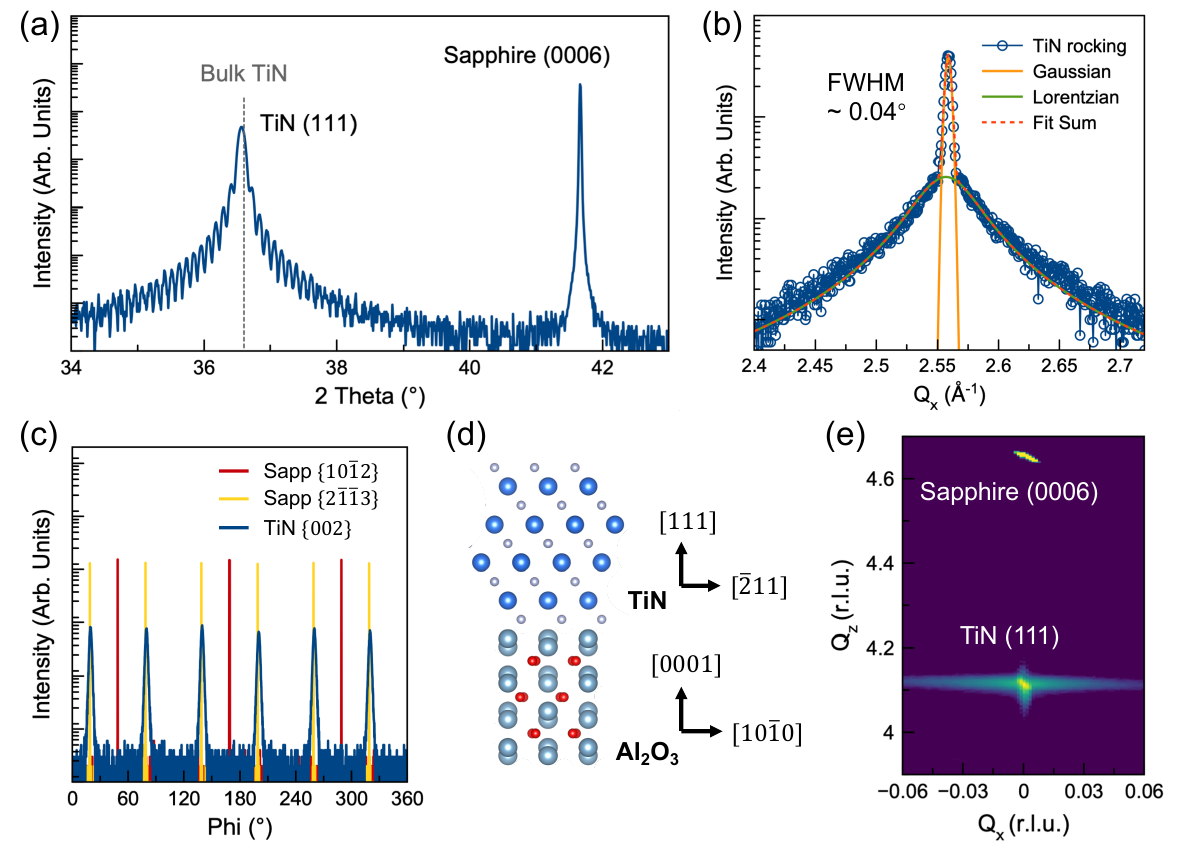}
\caption{(a) X-ray \textit{00l} line scan at TiN $(111)$- and sapphire $(0006)$-diffraction conditions. (b) Rocking curve scan at TiN $(111)$-diffraction condition. Subsequent analysis was performed by fitting the intensity using the pseudo-\textit{Voigt} function. (c) $\phi$ scans of TiN/Sapphire heterostructures to extract the epitaxial relationship as illustrated in (d). (e) On-axis reciprocal space mapping revealing a significantly broadened TiN peak in the in-plane direction.}
\label{fig:EpiTiNFig1} 
\end{figure*}

In this spirit, we focus herein on the epitaxial TiN thin films grown on $c$-cut sapphire substrates using a high-throughput RF-magnetron sputter system at an intermediate temperature of 300 $^{\circ}$C. A thorough material characterization was performed on the TiN/sapphire heterostructures. From X-ray-diffraction (XRD) and transmission-electron-microscopy (TEM) studies, the TiN/sapphire heterostructures were revealed to be highly crystalline and epitaxial, although detailed structural analysis suggested the pervasive existence of twist domains in the films. Chemical analysis including X-ray-photoelectron spectroscopy (XPS) and energy-dispersive X-ray-spectroscopy (EDX) mapping confirmed the TiN films to be stoichiometric. Following comprehensive material characterization, TiN/sapphire superconducting coplanar waveguide (SCPW) microwave resonators were fabricated using both dry- and wet-etch techniques. Cryogenic tests revealed that the highest internal quality factor $Q_i$ of the epitaxial TiN/sapphire resonators can reach $3.3\times 10^6$ in the single-photon regime and greater than $1\times 10^7$ in the high-power regime. Averaged single-photon quality factors greater than $1.4 \times 10^6$ can be repeatably achieved using both dry- and wet-etch techniques. The quality factors are comparable to the previously reported best values for TiN~\cite{Calusine2018,Ohya2014} and other material systems such as aluminum \cite{Megrant2012} and niobium \cite{Potocnik2020,Altoe2020}. This work lays the foundation for high-coherence superconducting quantum circuits and detailed decoherence studies on superconducting quantum devices. 

\section{Results \& Discussion}

\subsection{Structural Characterization}

100-nm-thick TiN thin films were deposited using a RF-magnetron sputtering system at an intermediate deposition temperature of 300~$^{\circ}$C (see Appendix \ref{appendix:A}). To examine the epitaxial relationship and structural properties, we first performed a set of X-ray line scans using a high-resolution X-ray diffractometer (see Appendix \ref{appendix:A}). $\omega$-2$\theta$ line scans around the $(0006)$-diffraction condition of the sapphire substrates (Figure~\ref{fig:EpiTiNFig1}(a)) revealed that the TiN is epitaxially grown with the $[111]$\textsubscript{TiN} axis aligned with the sapphire $[0001]$\textsubscript{Sapp} axis. The peak position at the TiN $(111)$-diffraction condition is well matched with that expected for the bulk and stoichiometric TiN. This suggests a fully relaxed TiN thin film grown on a sapphire substrate, which is consistent with the lattice mismatch between the TiN and sapphire being as large as $\sim$8\%. In addition, The diffraction peak of TiN showed clear \textit{Laue} diffraction fringes, which also indicates a high degree of film crystallinity.

To quantify the crystallinity of the TiN samples, a rocking curve was measured at the TiN $(111)$-diffraction condition (Figure~\ref{fig:EpiTiNFig1}(b)). The rocking curve revealed a full width at half-maximum (FWHM) of 0.04$^{\circ}$. To the best of our knowledge, this is among the best reported sputtered TiN films and even among some of the TiN films grown via molecular-beam epitaxy \cite{Guo2019,Olson:2015,Smith2018}. Low-temperature deposition of high-quality epitaxial TiN thin films was reported, but was done with a heavily customized sputtering system to modify the magnetic flux lines \cite{Petrov1992,Smith2018,Smith2020}. As such, using a low-thermal-budget deposition process with a readily used commercial deposition system presents opportunities for material investigation, process optimization, and device integration in future works. In addition to the high degree of crystallinity, we also observed that the rocking curve is composed of two features, namely the signal consists of one sharp peak that can be fitted with the Gaussian function and one broad peak that can be fitted with the Lorentzian function (Figure~\ref{fig:EpiTiNFig1}(b)). The Gaussian part of the rocking curve is attributed to the coherent diffraction of the TiN lattice, while the Lorentzian part is a result of the diffuse scattering from lattice imperfections, such as structural domains and defect clusters, which perturb long-range ordering \cite{Moram2009,Kaganer2009,Metzger1998}. To gain more quantitative insight into such a diffusive part, the peak intensity $I_{\text{rocking}}$ was conventionally fitted with a pseudo-Voigt function defined as $I_{\text{rocking}}=\eta I_G + (1-\eta) I_L$ $(0 \leq \eta \leq 1)$, where $I_G$ and $I_L$ are the Gaussian and Lorentzian function, respectively \cite{Kobayashi1999,Liu2008}. The Lorentzian function is given by $I_L = I_0 \frac{\pi\Gamma}{Q^2+\Gamma^2}$, and the correlation length $\xi = \Gamma^{-1}$ of the diffusive scattering part was extracted. It was found that the correlation length $\xi$ is on the order of 20$-$30 nm, and is likely to have originated from the microscopic domains in the films. We will further address the microscopic structures and the length scales of such domains in the following subsection with TEM studies. In all, the X-ray line scans and rocking curves all confirmed that the TiN thin films are of a high degree of crystallinity.

The in-plane epitaxial relations between TiN and sapphire substrates were examined by performing a set of off-axis $\phi$-scans (Figure~\ref{fig:EpiTiNFig1}(c)). The TiN $\{002\}$-diffraction peak series showed sixfold symmetry with each diffraction peak separated by 60$^{\circ}$. Since TiN has cubic symmetry, one should only observe threefold rotational symmetry if viewed along the $[111]$ direction. The presence of the six diffraction peaks thus indicates that there is an additional type of twist domain in the material where the domains are rotated along the substrate norm by 60$^{\circ}$. In addition, the six diffraction peaks have almost the same intensity, suggesting that there is no preference for the material to form certain types of domains. By examining the relative peak positions of the TiN $\{002\}$-diffraction peak series and the sapphire $\{10\bar{1}2\}$-, $\{2\bar{1}\bar{1}3\}$-diffraction peak series, we can thus draw a conclusion that the epitaxial relation between TiN and sapphire is given by $[111]$\textsubscript{TiN} $\parallel$ $[0001]$\textsubscript{Sapp} and $[\bar{2}11]$\textsubscript{TiN} $\parallel$ $[10\bar{1}0]$\textsubscript{Sapp} (Figure~\ref{fig:EpiTiNFig1}(d)). Such findings are consistent with previously reported results for TiN grown on \textit{c}-cut sapphire substrates \cite{Gupta2019}.

On-axis reciprocal space mapping (RSM) on the TiN/sapphire heterostructure was also performed to extract more information on the material microstructures (Figure~\ref{fig:EpiTiNFig1}(e)). The results revealed that there is significant peak broadening along the in-plane direction compared with the out-of-plane direction. We only show a two-dimensional cut aligned with the$[2\bar{1}\bar{1}0]$ axis of the sapphire substrates, while the scans along other in-plane axes showed no significant differences in terms of peak shapes. The peak broadening in both rocking curves and RSMs suggest that the lattice coherence along the in-plane direction is significantly worse compared to the out-of-plane direction. There are several possibilities that could give rise to the lateral peak broadening. First, the twist domains of the TiN could tilt away from the substrate norm and generate smeared diffraction patterns around the main diffraction peak. Such tilted domains are commonly seen in other nitride systems \cite{Kobayashi1999,Moram2009,Liu2008,Kobayashi1999}, and will be discussed in the next subsection. Second, high density of threading dislocations in the film with the non-zero out-of-plane component in the \textit{Burgers} vector could also give rise to such peak broadening. The exact peak profiles are largely dependent on the defect structures, which have been comprehensively studied in the III--V nitride systems \cite{Barchuk2010,Kaganer2005,Barchuk2018,Romanitan2017}. Third, the reduced domain sizes in the lateral direction, together with the high density of domain boundaries, could also disrupt the long-range lattice correlation, giving rise to diffraction-peak broadening. To gain more microscopic insights, we conducted TEM studies on the epitaxial TiN films.

\subsection{Microstructures and Domains}

\begin{figure} [t]
\includegraphics[width=8.6cm]{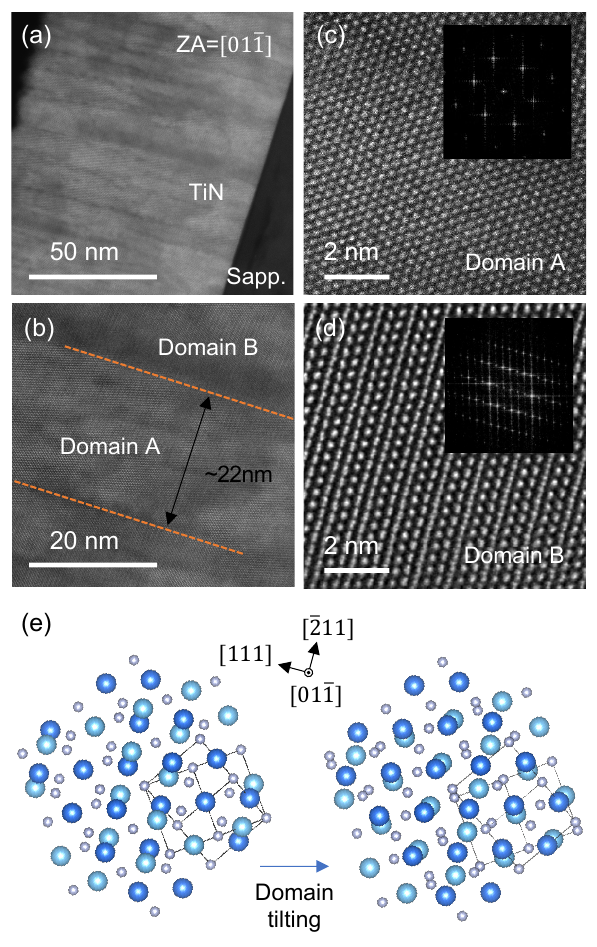}
\caption{(a) Low-magnification HAADF images of TiN/Sapphire heterostructures. (b) Zoomed-in image to show two domain structures in the TiN films. HAADF images of the two domain regions under high magnification are given, where one shows (c) a single lattice structure, while the other shows (d) a \enquote{superlattice} structure. (e) Schematic illustration of the formation of the \enquote{superlattice} structures. Upper panel shows the ideal superposition of two twist domains and lower panel illustrates the situation when the two twist domains are tilted. Briefly, as the two twist domains (light and dark blue) are slightly tilted away from the substrate norm, there is a systematic displacement of atomic columns on the projection plane, corresponding to the observed \enquote{superlattice} structures. }
\label{fig:EpiTiNFig2} 
\end{figure}

High-angle annular dark-field (HAADF) images were taken on the TEM/sapphire heterostructures with the zone axis aligned with $[01\bar{1}]$\textsubscript{TiN} (see Appendix~\ref{appendix:A}). Low-magnification HAADF images (Figure~\ref{fig:EpiTiNFig2}(a)) revealed a sharp epitaxial boundary and the existence of columnar-like domains elongated along the substrate norm. Based on previous X-ray studies, we argue that the columnar domains are mainly the twist domains. That is to say, the $[111]$\textsubscript{TiN} axes of the domains are all aligned with the $[0001]$\textsubscript{Sapp} axis (or mildly tilted away from the $[0001]$\textsubscript{Sapp} axis), while the in-plane $[01\bar{1}]$\textsubscript{TiN} axis of each domain is offset by 60$^\circ$. In a zoomed-in image (Figure~\ref{fig:EpiTiNFig2}(b)), one can clearly distinguish the different atomic arrangements corresponding to different domain regions. The region labeled \textit{Domain A} only contains one type of lattice structure consistent with the crystal structure of TiN, while the region labeled \textit{Domain B} contains alternating brighter atomic columns and yields a seemingly \enquote{superlattice} structure. The formation of the \enquote{superlattice} structure in the \textit{Domain B} region is due to the existence of more than one type of twist domain along the zone axis (to be addressed further in the following section). At the intersection of the two domains, dark stripes exist in the TEM image due to the presence of domain boundaries and high density of dislocations \cite{Kobayashi1999,Liu2008}. The long-range correlation in the in-plane direction of the TiN lattice was disrupted by the high density of incoherent domain boundaries, resulting in diffuse scattering featured in the rocking curves and RSM studies. If we measure the lateral dimension of a single twist domain, the length scale is on the order of 20 nm, in good agreement with the correlation length observed in the X-ray rocking curve analysis. Similar characterization consistency on domain dimensions using varied techniques was also reported in other nitride systems \cite{Metzger1998}.

To further confirm and analyze the twist domain structures, high-magnification images were taken at two distinct domains (Figures~\ref{fig:EpiTiNFig2}(c) and (d)). It was revealed that for the regions in \textit{Domain A} (Figure~\ref{fig:EpiTiNFig2}(c)), a single-lattice structure was observed, corresponding to the TiN lattice projected along the $[01\bar{1}]$\textsubscript{TiN} direction [see the fast \textit{Fourier} transform, Figure~\ref{fig:EpiTiNFig2}(c), inset]. This also suggests that there is only one type of domain in this region along the zone axis. For the regions in \textit{Domain B} (Figure~\ref{fig:EpiTiNFig2}(d)), however, we observed a \enquote{superlattice} structure that does not match the expected TiN structures. The \enquote{superlattice} structure is basically an alternation of two clear rows of atoms and one opaque row of atoms (note that we use \textit{row} to refer to a chain of atoms perpendicular to the zone axis, and \textit{column} to refer to a chain of atoms in parallel to the zone axis). As a result, the fast \textit{Fourier} transform of this regime gave additional high-order diffraction spots (Figure~\ref{fig:EpiTiNFig2}(d), inset]. As such, we propose that the \enquote{superlattice} structure is essentially a result of two different tiled twist domains that are superpositioned along the zone axis. As given in the schematics (Figure~\ref{fig:EpiTiNFig2}(e)), the two types of twist domains are illustrated using different colors (light and dark blue) with coordinates purposely rotated to align with the orientations in the TEM images. If the two twist domains both have their $[111]$\textsubscript{TiN} axis perfectly aligned with the substrate norm, the projection along the zone axis should be two opaque rows of atoms plus one clear row of atoms [as given in Figure~\ref{fig:EpiTiNFig2}(e), upper panel], which is inconsistent with our TEM results. However, if the domains are slightly tilted away from the substrate norm, the tilts will manifest as a small but systematic displacement of atoms on the projection plane (Figure~\ref{fig:EpiTiNFig2}(e), lower panel]. Such systematic displacement results in a new lattice structure on the projection plane with one row of \textit{zigzag} atoms (opaque rows in the TEM image) and two rows of mildly displaced atoms (clear rows in the TEM image). This scenario is in good agreement with the TEM findings, which confirms the existence of tilted twist domains. Thus, the TEM studies revealed the existence of domains, which are further identified to be tilted twist domains with a lateral dimension on the order of 20~nm.

\subsection{Chemical Characterization}

\begin{figure} [b]
\includegraphics[width=8.6cm]{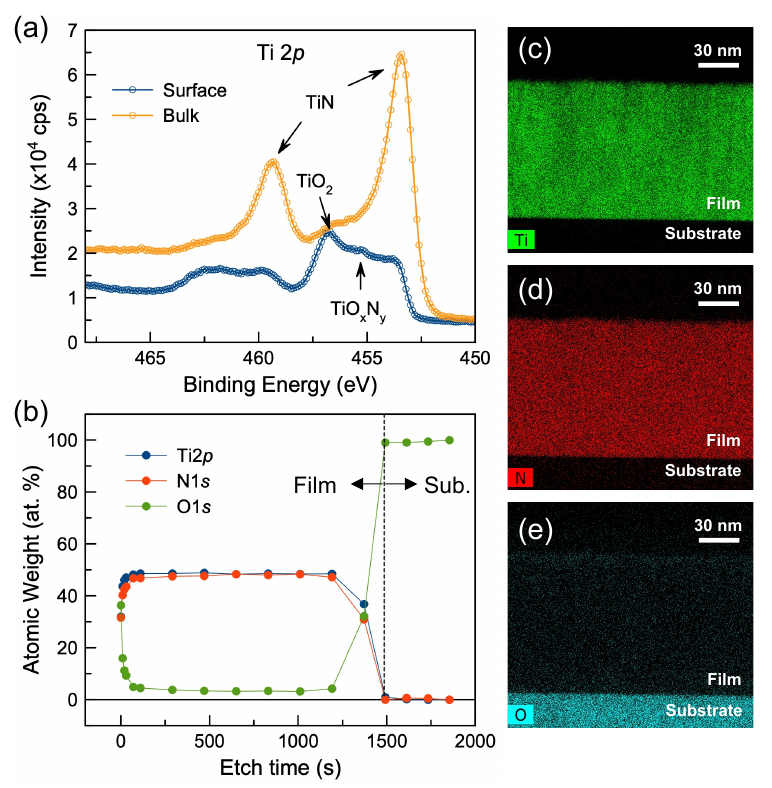}
\caption{(a) Ti 2\textit{p} spectra taken at the film surface and within the bulk part of the film. (b) XPS depth profile in units of atomic percentage. It can be seen that the film is essentially stoichiometric across the entire thickness except for the interfacial regimes. (c--e) EDX spectroscopy images taken during TEM characterization also revealed homogeneous distribution of elements and sharp interfaces.}
\label{fig:EpiTiNFig3} 
\end{figure}

After establishing a thorough understanding of the structural properties of the epitaxial TiN films, we continued to examine the chemical properties of the heterostructures. XPS analysis with \textit{in situ} ion milling was first performed to study the film chemistry (see Appendix~\ref{appendix:A}). As given by the XPS spectra (Figure~\ref{fig:EpiTiNFig3}(a)), the titanium 2\textit{p} spectra on the surface of the film clearly exhibit two sets of doublet peaks. One set corresponds to the Ti 2\textit{p}$_{3/2}$ and Ti 2\textit{p}$_{1/2}$ doublets of TiN, and another set corresponds to the same doublets, but with a stronger binding energy from TiO$_2$. In addition to these two sets of clearly resolved doublets, a third set is also visible in between the two sets, very likely having originated from the titanium oxynitrides formulated by TiO$_x$N$_y$ (x and y indicate the unclear stoichiometry of such compound). Thus, the surface of the TiN film is highly oxidized, which is consistent with previous findings \cite{Chang2019,Smith2020}. After milling the film for a sufficient amount of time, the spectra evolve where only the doublets from TiN dominate.

By integrating the peak area of the corresponding elements during ion milling, we plotted the XPS depth profile of the TiN/sapphire heterostructures (Figure~\ref{fig:EpiTiNFig3}(b)). As given in the figure, the atomic ratios of titanium and nitrogen are essentially equal to unity throughout the film, except for the interfacial regions. Thus, within the experimental resolution, we can conclude that the sputtered TiN films on sapphire substrates are stoichiometric. Near the surface of the film, a sharp change in film chemistry was observed due to the surface oxidation layer. While at film/substrate interface, the relatively wide interfacial region was due to the different milling rate of the sapphire substrate and the TiN films. It is worth noting that there is roughly 3--5 at.\% of oxygen content throughout the TiN films, and the existence of oxygen is very likely due to the relatively high base pressure of the sputter chamber ($>5 \times 10^{-5}$ Pa) before deposition. EDX studies were also performed during TEM studies. Three exemplary elements including titanium, nitrogen, and oxygen are given (Figure~\ref{fig:EpiTiNFig3}(c--e)). It was found that the film is highly homogeneous in terms of chemical composition across the film thickness, and there exists a sharp transition at the film/substrate interface, which is consistent with previously reported results \cite{Rasic2017}. In all, a comprehensive chemical analysis was applied on the epitaxial TiN films grown on sapphire substrates using both XPS and TEM techniques. The results showed that the TiN films are stoichiometric, in agreement with the XRD studies. The TiN films are oxidized at the surface, forming an ultra-thin layer of complex titanium oxynitride compound.

\subsection{Transport Properties}

\begin{figure} [t]
\includegraphics[width=8.6cm]{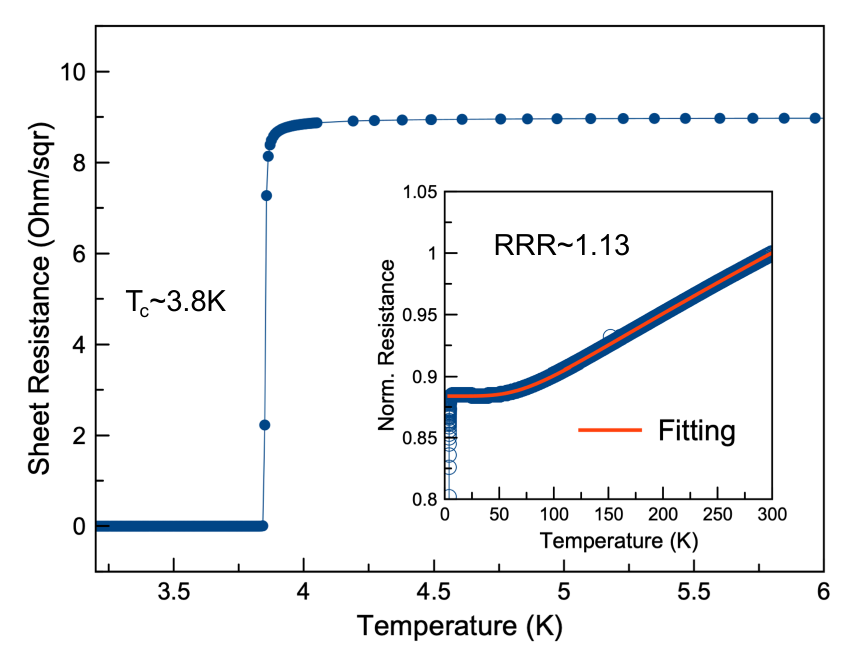}
\caption{Temperature-dependent sheet resistance of TiN/Sapphire heterostructure. Here, we observe a superconducting transition temperature of approximately 3.8~K. Normalized resistance fitted with \textit{Bloch-Grüneisen} formula in the high-temperature region is given in the inset.}
\label{fig:EpiTiNFig4} 
\end{figure}

\begin{figure*} [t]
\includegraphics[width=17.2cm]{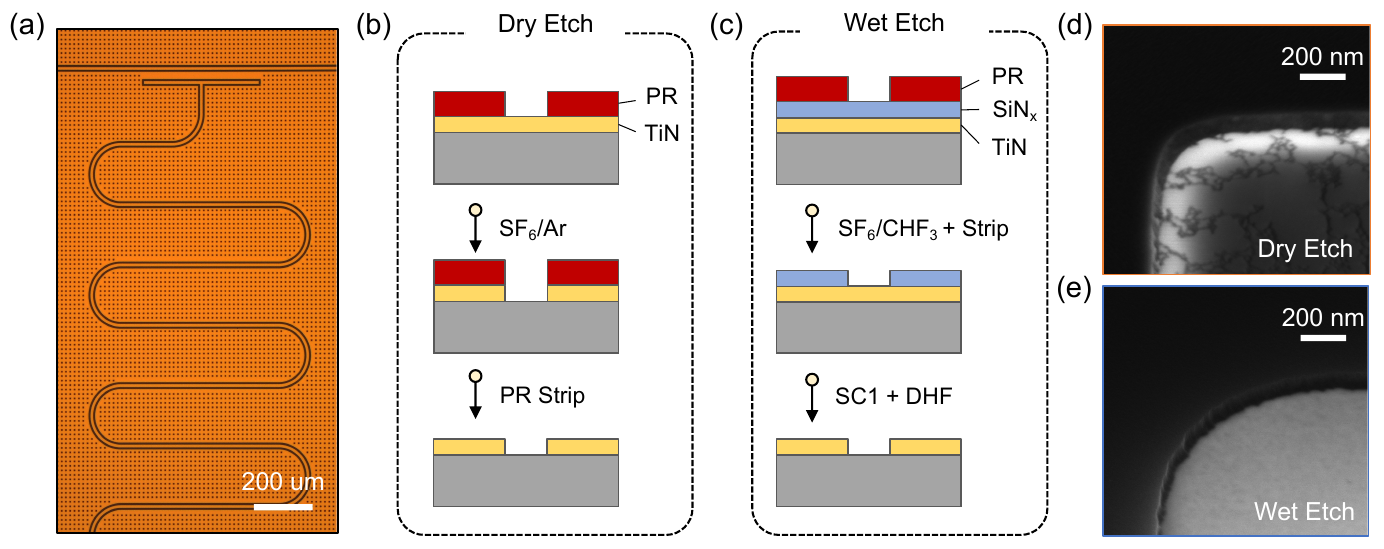}
\caption{(a) The optical image of one of the ten resonators in the device. The meandered half-wavelength CPW resonator is capacitively coupled to the readout line. Process flows of (b) dry- and (c) wet-etch techniques are given. (d,e) SEM images of the edge profiles of the fabricated devices. The dry-etch technique yields sharper edges, while there are signs of substrate surface damages or contamination; the wet-etch technique yields clean substrate surfaces, while the edges are slightly curvy. }
\label{fig:EpiTiNFig5} 
\end{figure*}

Transport properties of the epitaxial TiN/sapphire heterostructures were examined (see Appendix~\ref{appendix:A}) and the results given (Figure~\ref{fig:EpiTiNFig4}). A superconducting transition temperature ($T_c$) of $\sim$3.8 K was observed. Since the critical temperature $T_c$ is highly dependent on the deposition temperature, our observation is consistent with previous studies on epitaxial TiN films grown via magnetron sputtering and molecular-beam epitaxy at similar temperatures \cite{Krockenberger2012,Smith2020,Olson:2015}. A full temperature range of normalized resistance is also provided (Figure~\ref{fig:EpiTiNFig4}, inset). At higher temperatures, the TiN films showed clear metallic behavior and were fitted using $\rho = \rho_0 + A(\frac{T}{\Theta_D})^n \int_{0}^{\Theta_D/T} \frac{t^n}{(e^t-1)(1-e^{-t})} dt$. The total resistance is the sum of the residual resistance $\rho_0$, and the temperature-dependent part is given by the \textit{Bloch-Grüneisen} formula with $n=5$, which is a typical value for metallic films. The fitting describes the scattering events originated from electron-phonon interaction and yields a \textit{Debye} temperature $\Theta_D\sim$455 K, in reasonably good agreement with reported values for TiN thin films \cite{Saveskul2019,Postolova2017}. At lower temperatures, the resistance saturates to the residual resistance and a residual-resistance ratio ($RRR = \rho_{300 K}/\rho_{4 K}$) was extracted to be $\sim$1.13. Compared with the reported $RRR$ (in the range 2--7) of high-quality single-crystalline TiN films \cite{Krockenberger2012,Saveskul2019,Richardson2020}, the $RRR$ of this work is relatively small, suggesting that there is a significant number of lattice defects serving as scattering centers. Given that both XRD and TEM analysis both point to the pervasive existence of domain boundaries nearly parallel to the substrate norm (perpendicular to the electron transport directions), these domain boundaries could be the dominant source of electron scattering.

\subsection{Device Fabrication: Coplanar-waveguide Microwave Resonators}

We then proceeded to fabricate the TiN/sapphire heterostructures into coplanar-waveguide (CPW) resonators to examine the microwave properties at low temperatures. The CPW microwave resonators have a characteristic impedance of 50~$\Omega$, featuring a center-line width of 10~$\mu$m and a center-to-ground gap of 6~$\mu$m. Ten half-wavelength hanger-type resonators were capacitively coupled to a 50~$\Omega$ CPW for transmission measurements, and the ground plane perforated with $5\times 5$~$\mu$m$^2$-sized square openings for flux trapping (Figure~\ref{fig:EpiTiNFig5}(a)). To pattern the devices, both dry- and wet-etch techniques were developed. The dry-etch technique is relatively straightforward in terms of fabrication processes (see Appendix~\ref{appendix:B}). Nevertheless, a potential drawback of this process is that the etchant gases will inevitably reach and damage the substrate surface, which is undesirable if the subsequent processes have stringent requirements on the substrate-surface qualities (\textit{e.g.}, the epitaxial growth of another layer of material). To protect the substrate surfaces for the following fabrication steps, we introduced a hardmask-based wet-etch technique in which a silicon nitride (SiN$_x$) layer is used to mask the TiN features in the SC-1 etchants (a mixture of ammonium hydroxide and hydrogen peroxide solutions; see Appendix~\ref{appendix:B}). Here, we briefly describe the fabrication details of the two approaches.

First, for the dry-etch technique (Figure~\ref{fig:EpiTiNFig5}(a)), a single layer of photoresist (PR) was first coated on the TiN/sapphire heterostructures, followed by lithographic patterning. The device was then etched by a mixture of SF$_6$/Ar etchants in an inductively coupled plasma system (ICP; see Appendix~\ref{appendix:B}). After the dry etching, the protective photoresist was stripped in the organic solutions. From the scanning electron microscopy (SEM) images (Figure~\ref{fig:EpiTiNFig5}(c)), the ICP etching process yields sharp and clean TiN edges. Although the etching recipe in principle does not attack sapphire substrates, some meandering features can still be observed on the sapphire surfaces, which could be indicative of surface contamination or damage after the exposure to etchant gases. For the wet-etch technique (Figure~\ref{fig:EpiTiNFig5}(b)), a 100-nm-thick SiN$_x$ hardmask was deposited on the TiN/sapphire heterostructures via a plasma-enhanced chemical-vapor-deposition system (PECVD; see Appendix~\ref{appendix:B}). The material stack was coated with photoresist followed by lithographic patterning, and then dry-etched using the ICP system to form the device structures. After photoresist stripping, the hardmask-patterned TiN/sapphire heterostructures were immersed in the SC-1 solution heated to 60~$^{\circ}$C. The SC-1 solution etches TiN films and removes the residual organic contamination (see Appendix~\ref{appendix:B}). The bilayer stack was eventually immersed in a diluted hydrofluoric solution (DHF) to strip off the SiN$_x$ mask layer. As demonstrated in the SEM figure (Figure~\ref{fig:EpiTiNFig5}(d)), the wet-etch technique yields both a clean sapphire surface and a TiN top surface. Nevertheless, the profile edges of the wet-etched TiN showed curvy features, which might be a result of the additional pattern-transfer step through the SiN$_x$ hardmask layer and is worth further optimization.

\subsection{Cryogenic Microwave Studies}

\begin{figure} [t]
\includegraphics[width=8.6cm]{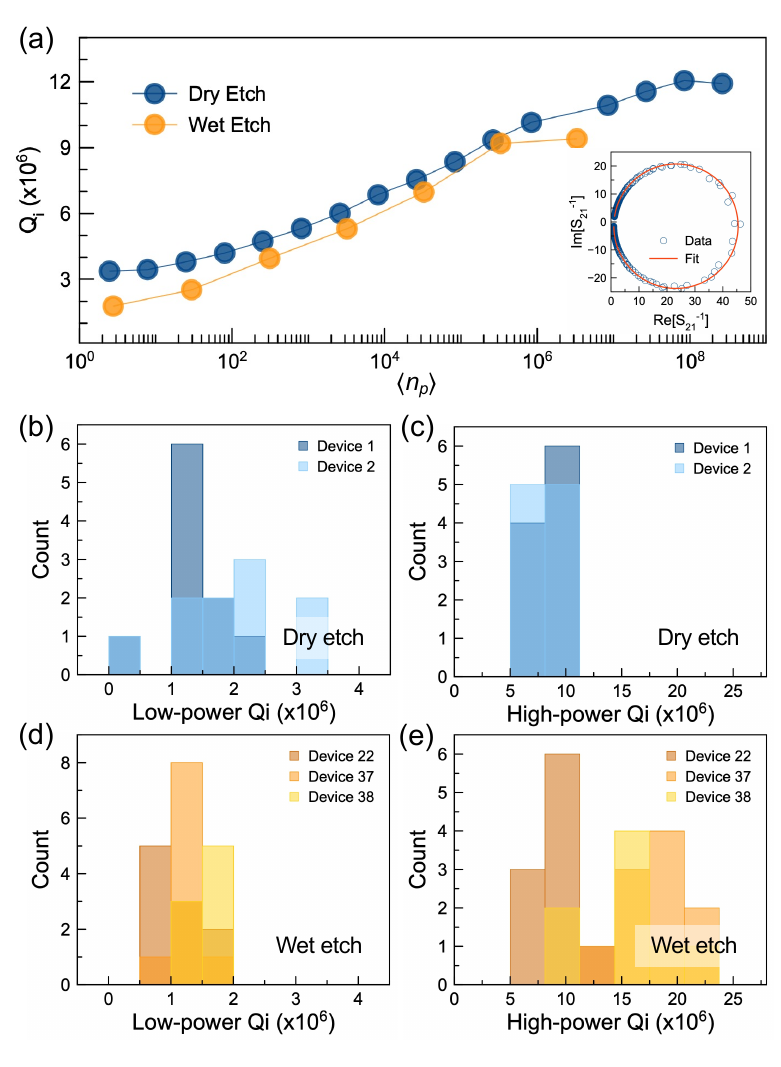}
\caption{(a) Photon-number-dependent quality factors of the best resonators are given for devices fabricated using different processes. Inset shows a typical numerical fitting to extract the internal quality factor. The quality factors of the TiN resonators are repeatable in different device batches for both dry-etch technique (b,c) and wet-etch technique (d,e). Note that each device contains 10 resonators and the statistical results of the quality factors at the lowest and the highest input power were summarized for each device, respectively. }
\label{fig:EpiTiNFig6} 
\end{figure}

We further characterized the microwave properties of the TiN/sapphire heterostructures using the as-fabricated resonator samples. The samples were rinsed in the buffered-oxide etch (BOE) for 5 min to remove the surface oxides before packaging into the aluminum sample boxes. We did not perform fast loading for the devices, and the time interval between the BOE rinse and the packaging ranges from several hours to several days for different device batches. The characterization was performed at $\sim$10~mK using the transmission scattering parameter ($S_{21}$) of a microwave feedline capacitively coupled to 10 resonators with characteristic resonance frequencies around 6 GHz (see Appendix~\ref{appendix:C} for details of the cryogenic setup). The internal quality factors ($Q_i$) of the resonators were extracted by numerically fitting the inverse $S_{21}$ spectra using the formula $S_{21}^{-1}(f) = 1 + \frac{Q_i}{Q_c^*}e^{i\phi} \frac{1}{1+2 i Q_i \frac{f-f_0}{f_0}}$, where $f_0$ is the resonance frequency, $Q_i = (Q^{-1}-Q_c^{-1})^{-1}$ is the internal quality factor of the resonator, $Q_c^*=Q_c (Z_0/|Z|)$ is the rescaled coupling quality factor, $e^\phi$ is an additional imaginary factor to describe impedance mismatch, $Z_0$ is the characteristic impedance of the input/output ports, and $Z = |Z|e^{i\phi}$ is half the inverse sum of the environment impedance on the input and output sides of the resonators (Figure~\ref{fig:EpiTiNFig6}, inset) \cite{Wang2020,Megrant2012}. 

To extract the resonator loss at the single-photon regime, the internal quality factors of TiN/sapphire microwave resonators were measured as a function of input microwave power (Figure~\ref{fig:EpiTiNFig6}(a)). The input power $P_{\text{in}}$ was converted to photon numbers in the resonator using the relation $\langle n_p \rangle =\frac{2Q^2P_{\text{in}}}{\hbar \omega_0^2 Q_c}$, where $Q$ is the total quality factor, $P_{\text{in}}$ is the input microwave power at the device, $\omega_0 = 2\pi f_0$ is the angular resonance frequency of the resonator, and $Q_c$ is the coupling quality factor extracted from the fitting. At the single-photon limit, the best device in the TiN/sapphire resonators fabricated using the dry-etch technique showed an internal quality factor $Q_i = 3.3\times 10^6$, which is among one of the best values reported for TiN systems~\cite{Ohya2014,Calusine2018}. The averaged $Q_i$ values of the dry-etched TiN resonators is $2.0 (\pm 0.73) \times 10^6$, and such high $Q_i$ can be repeated in different experimental batches (Figure~\ref{fig:EpiTiNFig6}(b,c)). For the TiN/sapphire resonators fabricated using the wet-etch technique, we have also witnessed a reasonably consistent results across different wafer batches. The best $Q_i$ was revealed to be $1.8\times 10^6$, slightly lower compared with the dry-etching approach, while the averaged $Q_i$ of wet-etched resonators is $1.4 (\pm 0.34) \times 10^6$ (Figure~\ref{fig:EpiTiNFig6}(d,e)). The relatively lower quality of the wet-etched resonators is subject to further investigation and improvement in the fabrication processes. 

In addition, the observed microwave-power dependence of the resonator quality factors indicate that the system is coupled to two-level systems (TLSs), which dominate the energy loss at the single-photon regime~\cite{Phillips1987, Martinis2005}. To quantitatively understand the impact of TLSs, we estimate the TLS quality factor $Q_{\text{TLS}}$ by using $1/Q_{\text{TLS}} \approx 1/Q_{i,\text{LP}} - 1/Q_{i,\text{HP}}$, where $Q_{i,\text{LP}}$ and $Q_{i,\text{HP}}$ are the quality factors at low power and high power, respectively. $Q_{\text{TLS}}$ is extracted to be as high as $4.2\times10^6$ for the best devices and averaged at $3.0 (\pm0.94)\times10^6$. The origin of the phenomenological TLS can be attributed to a number of microscopic mechanisms~\cite{Calusine2018,Sage2011,Muller2019,McRae2020}. For instance, the TLS could be located at the oxidation layer on the top and sidewalls of the TiN films, or the imperfections at the film/substrate interfaces. As reported in the niobium resonators~\cite{Altoe2020, Verjauw2021}, rapid loading (less than 20 min) after oxide etch could have a drastic impact on device performance and TLS contribution. Although 5 min of etching in the BOE solution was applied on the TiN resonator devices, since we have yet performed fast loading, the oxidized compounds could rapidly form during the time interval between device cleaning and eventual cryogenic tests. In addition, the film/substrate interfaces are known to have the largest electric-field energy participation \cite{Wang2015, Gambetta2017}; thus, lattice imperfections extended to the interface regions such as thread dislocations and grain boundaries could potentially harm the device quality. In all, the epitaxial TiN/sapphire systems revealed low microwave loss represented by resonator internal quality factors up to $3.3\times 10^6$ at the single-photon regime and $> 1\times 10^7$ at the high-power regime. The best and averaged loss tangent $tan\delta=\frac{1}{Q_{i,\text{LP}}}$ of the epitaxial TiN/sapphire resonators are $\sim3.0\times10^{-7}$ and $\sim7.1\times10^{-7}$, comparable to the previously reported best values for superconducting resonators~\cite{McRae2020, Altoe2020}.

\section{Conclusions}

In this work, epitaxial and highly crystalline TiN/Sapphire heterostructures were sputtered at an intermediate temperature of 300 $^{\circ}C$. A set of systematic studies was performed to investigate the structural, chemical, electrical, and microwave properties. The epitaxial TiN films are revealed to be stoichiometric and the film crystallinities are quantified by a rocking-curve FWHM of 0.04$^{\circ}$. Detailed characterization via XRD and TEM reveal that there exists a high density of slightly tilted twist domains in which the lateral domain sizes are on the order of tens of nanometers. Transport studies reveal a $T_c$ of 3.8~K, and the low-temperature conductivity is mainly limited by defect scattering, consistent with the high density of twist domains as suggested by the structural characterizations. Both dry- and wet-etch approaches were developed for microwave resonator fabrication, and the cryogenic measurement at the single-photon regime reveals a best $Q_i$ value as high as $3.3\times 10^6$. For different fabrication approaches, the resonator $Q_i$ can repeatably reach an averaged value of $> 1.4\times 10^6$. The low deposition thermal budget, high epitaxial-film quality, and low microwave losses make the epitaxial TiN/sapphire heterostructure an ideal platform for future decoherence studies and high-coherence superconducting quantum devices.

\begin{acknowledgments}
   We thank all members of Alibaba Quantum Laboratory for the fruitful scientific input and support on experiments and data interpretation. We thank the Westlake Center for Micro/Nano Fabrication, Westlake University, for assistance with device patterning and discussions. We thank the Instrumentation and Service Center for Physical Sciences, Westlake University, for material characterization services and data analysis. 
\end{acknowledgments}

\bibliography{EpiTiN_Ref}

\clearpage

\begin{appendix}

\section{Thin-film characterization}
\label{appendix:A}

\textbf{Thin-film deposition.} The sputter system used for this study is a CS-200 sputter system (\textit{ULVAC Inc.}). The deposition chamber is equipped with a turbo-molecular pump that reaches a base-pressure of $5 \times 10^{-5}$ Pa, and four 4-inch sputter guns. Titanium nitride deposition was performed at the RF sputter gun using a 99.999\% titanium target. The deposition power is set to be 600W and the gas mixture (Ar:N$_2$=4:1) used for reactive sputtering is held at 0.5 Pa. The substrate-to-target distance was held at 110 mm throughout the sputtering. The deposition parameters are determined after a systematic optimization metrology, where the condition that yields the lowest TiN resistivity are used with a deposition rate of 0.045 nm/s.  Before the deposition, the as-received Sapphire substrates (single-side polished, \textit{Suzhou RDMICRO Co., Ltd.}) were first heated to 300$^\circ$C, held for 10 mins in vacuum, and then 5 mins in the Ar/N$_2$ gas mixture. After deposition, the samples were cooled and unloaded from the load-lock by venting with air.

\textbf{X-ray diffractometry (XRD).} The XRD studies including linescans, rocking curves, and phi-scans were performed on a high-resolution D8 ADVANCE X-ray diffraction system (\textit{Bruker}) with optics set up for epitaxial-film studies. For the on-axis RSM scans given in the main text, the $\phi$ angle of the sample was first aligned using one of the sapphire $\{2\bar{1}\bar{1}3\}$-diffraction peaks.  For another on-axis scan with a different $\phi$, the sample was aligned using one of the sapphire $\{10\bar{1}2\}$-diffraction peaks, which is essentially a rotation of the sample by 30$^{\circ}$ along the substrate norm.

\textbf{Transmission electron microscopy (TEM).} Cross section samples were prepared with the focused ion beam (FIB) technique on an FEI Helios 650 system. The high-resolution transmission electron microscopy (HRTEM) and the high-angle annular dark-field (HAADF) images were acquired using JEOL JEM-ARM200F electron microscope, which is equipped with a thermal field-emission gun and spherical aberration (Cs) corrector. The semi-convergence angle is around 23 mrad during HAADF imaging, and the collection angles of the detectors are 50 and 200 mrad. Energy-dispersive X-ray spectroscopy mapping was carried out on an FEI Talos F200X TEM at 200 kV equipped with a 4-inch column SDD Super-X detectors.

\textbf{X-ray photo-electron spectroscopy (XPS).} The XPS studies were performed on a ESCALAB Xi+ XPS system (\textit{Thermal Fisher}) at an incident angle of 60$^{\circ}$. The system was first calibrated with standard samples, and the selective milling between titanium, aluminum, oxygen, and nitrogen was confirmed to be minimal. For the XPS depth-profile studies, the milling area was set to be 2 mm by 2 mm while the analyzing area was concentric with the milling area and set to be 0.4 mm by 0.4 mm. High milling current of the $Ar^+$ was applied, and the beam energy was set to 2000 eV. The spectra were taken after every milling step in a 300 sec interval.

\textbf{Transport studies.} The transport studies were performed on a 4mm by 4mm TiN/Sapphire sample by wire-bonding (aluminum wires) the samples in a \textit{Van der Pauw} geometry. The samples were then loaded in a TeslatronPT-14T Cryofree superconducting magnet system (\textit{Oxford Instruments}). A constant current of 0.1 mA was supplied to the outer two contacts and the voltage drop between the inner two contacts was measured using a 2400-series source meter (\textit{Keithley Instruments}).

\section{Device fabrication}
\label{appendix:B}

\textbf{Plasma-enhanced chemical vapor deposition (PECVD).} The PECVD deposition of SiN$_x$ hardmask was performed in a Haasrode-C200A PECVD system (\textit{LEUVEN Instruments}). The process was done at 200 $^{\circ}$C in a gas mixture of N$_2$, SiH$_4$, and NH$_3$. The deposition pressure was maintained at 500 mtorr and a RF power set at 100 W.

\textbf{Inductively-coupled plasma (ICP).} The ICP process was performed for TiN dry etch and hardmask dry etch on a Haasrode-E200A ICP system (\textit{LEUVEN Instruments}) and a PlasmaPro100 ICP system (\textit{Oxford Instruments}), respectively. The process for TiN dry etch was done at 20 $^{\circ}$C in a gas mixture of SF$_6$ and Ar. Etching rate was calibrated independently and the etching time for the actual devices was set to be 60 sec. For SiN$_x$ hardmask dry etch, the process was done at 20 $^{\circ}$C in a gas mixture of SF$_6$ and CHF$_3$. The ICP power was set at 800 W and the bias power was set at 20 W.

\textbf{Wet-etch processes.} As discussed in the main text, the wet etching of TiN films were performed in the SC-1 solution. The SC-1 solution used was composed of six parts of de-ionized water, one part of ammonium hydroxide (29\% by weight of NH$_3$), and one part of 30\% hydrogen peroxide. During the wet etch, the trilayer stack was immersed in the SC-1 solution heated to 60 $^{\circ}$C for 6 mins with continuous stirring, rinsed in a 60 $^{\circ}$C de-ionized water afterwards, and then cleaned with de-ionized water using continuous flow. To strip off the hardmask, the patterned trilayer stack was immersed in the diluted hydrofluoric acid, and then cleaned with de-ionized water using continuous flow.

\section{Cryogenic measurement}
\label{appendix:C}

\begin{figure} [H]
\includegraphics[width=8.6cm]{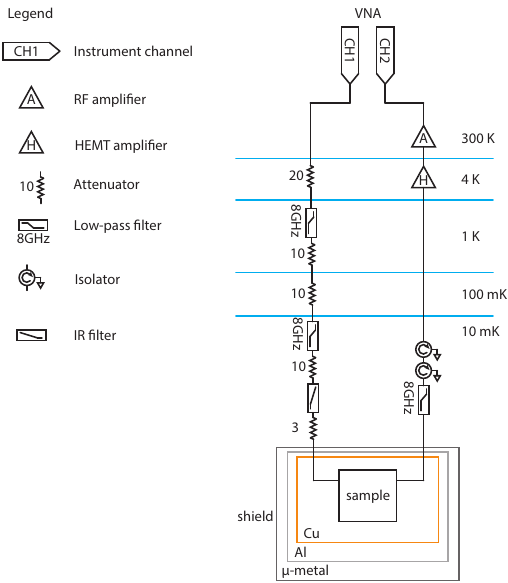}
\caption{The schematics of the cryogenic measurement setup for superconducting microwave resonators. The attenuation (in unit of dB) of attenuators and the cutoff frequencies of the low-pass filters are annotated beside the legends in the schematic correspondingly.}
\label{fig:EpiTiNFig9} 
\end{figure}

The cryogenic measurement setup for SCPW microwave resonators is illustrated (Figure~\ref{fig:EpiTiNFig9}). The dilution refrigerator system used for this study is a commercial LD400 system (\textit{Bluefors}). The microwave signal was generated by a E5071C ENA Vector Network Analyzer (VNA, \textit{Keysight Technologies}) and sent to the DUT at the base temperature ($\sim 10$ mK) through a series of cryogenic attenuators (\textit{XMA Inc.}), low-pass filters (F-30-8000-R, \textit{RLC Inc.}), and home-made infrared (IR) filters. The device under test was isolated from the environment by a layer of Cu-Al shield and a $\mu$-metal magnetic shield. The output signal from the DUT was sent through a low-pass filter (F-30-8000-R, \textit{RLC Inc.}) and a 2-stage isolator (LNF-ISISC4$\_$8A, \textit{LNF Inc.}) to reduce the possible noise and back reactions from the upper stages at higher temperatures. The output signal was then amplified using a high-electron-mobility transistor (HEMT) amplifier (LNC4$\_$8C, \textit{LNF Inc.}) mounted at the $4$ K stage, and sent back to the VNA after an additional amplifier (LNR4$\_$8C, \textit{LNF Inc.}) at room temperature.

\end{appendix}

\end{document}